\documentclass{aa}  % for the final two-column version

\usepackage{graphicx}

\newcommand{\micro}{\mbox{\usefont{U}{eur}{m}{n}\char22}}

\begin{document}

   \title{Mrk~421, Mrk~501, and 1ES~1426+428 at 100~GeV with the CELESTE Cherenkov Telescope}

%   \subtitle{The CELESTE Collaboration}
\titlerunning{Mrk~421, Mrk~501, and 1ES~1426+428 at 100~GeV}
\authorrunning{CELESTE}

   \author{
D.~A.~Smith\inst{1} % \thanksref{correspondents}
\and E.~Brion\inst{1}\thanks{{\emph Current address:} SAp, Commissariat \`a l'\'energie atomique, 91191 Gif-sur-Yvette, France}
\and R.~Britto\inst{2}
\and P.~Bruel\inst{3}
\and J.~Bussons Gordo\inst{2}
\and D.~Dumora\inst{1}
\and E.~Durand\inst{1}
\and P.~Eschstruth\inst{4}
\and P.~Espigat\inst{5}
\and J.~Holder\inst{4}
\and A.~Jacholkowska\inst{2}
\and J.~Lavalle\inst{2}
\and R.~Le Gallou\inst{1}
\and B.~Lott\inst{1}
\and H.~Manseri\inst{3}
\and F.~M\"unz\inst{5}
\and E.~Nuss\inst{2}
\and F.~Piron\inst{2}
\and R.~C.~Rannot\inst{1,6}
\and T.~Reposeur\inst{1}
\and T.~Sako\inst{3}
}

\institute{
Centre d'\'etudes nucl\'eaires de Bordeaux Gradignan - CENBG UMR 5797 CNRS/IN2P3 - Universit\'e Bordeaux~1, 
Chemin du Solarium - BP120 33175 Gradignan cedex, France
\and
Laboratoire de physique th\'eorique et astroparticules, Universit\'e Montpellier~2, 
34095, France (UMR 5207 CNRS/IN2P3)
\and
Laboratoire Leprince-Ringuet, \'Ecole Polytechnique, 
91128 Palaiseau, France (UMR 7638 CNRS/IN2P3)
\and
Laboratoire de l'acc\'el\'erateur lin\'eaire, Universit\'e Paris~11, 
91898 Orsay, France (UMR 8607 CNRS/IN2P3)
\and
Laboratoire d'astroparticule et cosmologie, Coll\`ege de France, 
75231 Paris, France (UMR 7164 CNRS/IN2P3)
\and
Astrophysical Sciences Division, Bhabha Atomic Research Centre,
Mumbai-400 085, India
}
   \offprints{smith@cenbg.in2p3.fr}

   \date{version: 19 August 2006 }

   \abstract{
We have measured the gamma-ray fluxes of the blazars \object{Mrk~421} and \object{Mrk~501} 
in the energy range between 50 and 350~GeV ($1.2$ to $8.3 \times 10^{25}$~Hz). 
The detector, called CELESTE, used first 40, then 53~heliostats of the former solar facility ``Th\'emis'' in the 
French Pyrenees to collect Cherenkov light generated in atmospheric particle cascades.
The signal from Mrk~421 is often strong. 
We compare its flux with previously published multi-wavelength studies and infer that we are straddling 
the high energy peak of the spectral energy distribution. The signal from Mrk~501 in 2000 was weak ($3.4~\sigma$). 
We obtain an upper limit on the flux from 1ES~1426+428 of less than half that of the Crab flux near 100~GeV. 
 The data ana\-ly\-sis 
and understanding of systematic biases have improved compared to previous work, increasing the detector's sensitivity.
   }

   \maketitle
%
%________________________________________________________________

\section{Introduction}

Measurements of high energy emission from active galactic nuclei (AGN) give insights into a variety of open problems,
such as the nature of the AGNs themselves, and the density and evolution of the extragalactic diffuse infrared
background. After GLAST is launched in 2007 (\cite{glast}), the large number of high galactic latitude GeV gamma-ray
sources to be seen will make AGNs become a background for searches for new classes of emitters~-- insufficient
understanding of high energy AGNs might just limit potential glimpses of a hidden Universe.
Several AGNs of the ``blazar'' class emit above 250~GeV and have been detected by atmospheric Cherenkov imaging
telescopes. They are called HBLs, for high energy BL Lac objects.
Of these, Mrk~421 (\cite{blaze,cat421,hegra421b}) and Mrk~501 (\cite{ir,cat501}) are the brightest . 

The Synchrotron Self Compton model (``SSC'') is consistent with much of the
multiwavelength blazar data. For discussion of these models as applied to the blazars
discussed in this paper, see for example (\cite{TMG}) or (\cite{KSK}). In its simplest variant, relativistic electrons
generate synchrotron photons with a $\nu F_\nu$ peak in the optical to X-ray range, and then up-scatter these same
photons to produce the peak seen in the GeV range. The inverse Compton peak position can constrain the
physical parameters of the model. 
Many, if not most, observed characteristics of blazars remain unexplained. Examples are the relation
between variations at different wavelengths, as highlighted by the ``orphan'' gamma ray flares observed
for 1ES1959+650 (\cite{kwaz1959}), and the role that proton acceleration might play.
Progress requires simultaneous measurements across the spectrum. 

The CELESTE telescope was one of the first instruments sensitive near the Compton peak for
these blazars, beyond the energy range of the rela\-ti\-ve\-ly large sample of EGRET blazars
(\cite{egretblazars,3rdcatalog}). Mrk~421 was studied with EGRET, but Mrk~501 was visible only during a flare (\cite{egret501}),
while 1ES 1426+428 was undetected by EGRET, although well-measured at TeV energies (e.g. \cite{cat1426,whipple1426}).
A key motivation for our experiment was to provide
blazar measurements in an unexplored energy window. 

%Until recently,
%the most distant blazar ($z = 0.129$) measured was 1ES~1426+428 (e.g. \cite{cat1426,whipple1426}). Using the similar
%distances of the two Markarians ($z \simeq 0.03$) and the relatively large distance to 1ES~1426+428, (\cite{ir}) were
%able to constrain the infrared background, via gamma-ray absorption by $e^+e^-$ pair production off of diffuse infrared
%radiation. High infrared densities, and/or a {\em very} hard GeV/TeV spectrum for 1ES~1426+428, lead to predictions of
%Crab-like intensities for 1ES~1426+428 at 100~GeV. Recent results from HESS seem to exclude these high densities
%(\cite{hessIR}), nevertheless one of our early motivations was to use CELESTE to see whether in fact 1ES~1426+428 is
%gamma-bright. We conclude that it is not, which confirms the trend indicated by those two articles.

This paper presents the first observations of three blazars below 100~GeV, by a solar heliostat array, and is
structured as follows: we describe the detector, the analysis method, and the data samples. We emphasize improvements
in our knowledge of the systematic uncertainties. We present light curves for the three blazars as well as the
integrated fluxes, and place the results in the context of SSC models. 
%__________________________________________________________________

\section{The CELESTE gamma-ray Telescope}

\begin{figure}
%\hspace*{-1.5cm}
%\centering
\resizebox{\hsize}{!}{\includegraphics[width=0.7\textwidth]{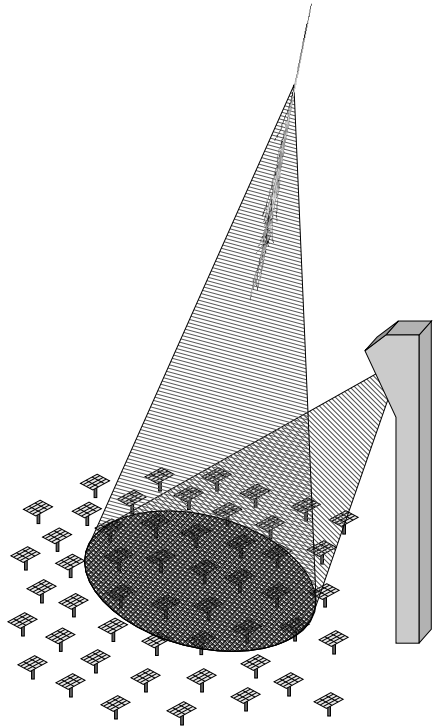}}
\caption{Experimental principle: Heliostats tracking a source reflect Cherenkov light from atmospheric particle
cascades to the secondary optics, photomultipliers, and acquisition electronics located high in the 100~meter tall
tower. 
\label{principle}}
\end{figure}

The CELESTE detector is located in the eastern French Pyrenees (N. $42.50^{\circ}$, E. $1.97^{\circ}$, altitude
$1650~\mathrm{m}$) and described in (\cite{bgnim,repoman}) and in (\cite{ourcrab}). Figure~\ref{principle} illustrates the
principle. CELESTE was dismantled in June, 2004.  

CELESTE detected the Crab nebula in 1998 using a solar heliostat array (\cite{Smith98}). STACEE, very similar to
CELESTE, reported the flux above 190~GeV (\cite{staceecrab}), followed by the flux above 60~GeV from CELESTE
(\cite{ourcrab}). STACEE recently measured the integral flux above 140~GeV from \object{Mrk~421} (\cite{stacee421})
and, assuming power law spectral indices extrapolated from higher energies, bracketed the 50 to 700~GeV differential
spectrum. A review of solar tower gamma-ray telescopes can be found in (\cite{Smith05}). Here we recall key features
of the CELESTE detector.

% aa-package/aadoc.pdf explains figures etc quite nicely

CELESTE used 53~heliostats (40 until 2001) of the Th\'emis former solar facility. The CAT Cherenkov imager ran on the same site
(\cite{cat421} and references therein), as did the \textsc{Themistocle} (\cite{themistocle}) and ASGAT (\cite{asgat}) Cherenkov sampling
arrays. Each heliostat has a $54~\mathrm{m}^2$ mirror on an alt-azimuth mount. Light from the heliostats was reflected to a secondary
optical system at the top of the $100$~m tower, where one photomultiplier assembly (PMT) viewed each heliostat. Heliostat control and data
acquisition systems were also high in the tower.

Data acquisition was triggered when the summed PMT pulse height for at least $M$ of the $N$ groups of heliostats exceeded
$4.5$~photoelectrons (``p.e.'') per heliostat. For the 40~heliostat array, $M=3$ of $N=5$ groups with 8~heliostats each was used most
often. For the 53 heliostat array there were $N=6$ trigger groups of between 6 and 8~heliostats each, and the trigger multiplicity was set
at $M=3, 4,$ or $5$ for different observations. The deadtime in the electronic delays depends on the rates, which depend on the threshold
settings, which are determined in part by the choice of $M$. We took pains to remove these biases (\cite{HMthesis}), in particular, with an
offline group multiplicity cut based on the scaler rates. The acceptance shown in figure~\ref{acceptc} has an offline multiplicity cut of 4
of 5, for illustration.

Each PMT signal was digitized at $1.06$~nanoseconds per sample, with about 3~digital counts (dc) per photoelectron. 100~samples centered
around the nominal Cherenkov pulse arrival time were recorded, providing event-by-event pedestal information that we used to determine the
night sky background light for each channel. A second 28~sample window contained a timing reference pulse. Scaler rates, PMT anode
currents, a GPS time stamp, and some meteorological information rounded out the data record.

\section{Monte Carlo Simulations} CELESTE's first publication, on the flux from the Crab nebula quoted a $\sim 30$~GeV threshold at the
trigger level, and 60~GeV after analysis (\cite{ourcrab}). We estimated the uncertainty on the energy scale to be less than 30~\%,
dominated by disagreement between predictions of Monte Carlo atmospheric cascade generators, and disagreement between estimated and
observed PMT illuminations induced by bright stars. 
Since then we have corrected some small errors in the KASCADE Monte Carlo (\cite{kaskade}) and the agreement is improved 
($\delta \epsilon_{MC} = 5$ to 10 \%).
The CORSIKA Monte Carlo atmospheric shower
generator (\cite{corsika}) best reproduces our experimental observables, and we use it for the results presented here.

Cherenkov photons from the shower generator are fed into a detailed simulation of our optics and electronics, to develop our data analysis
methods and cuts, as well as to calculate the energy dependent effective area $A(E)$. Here we describe improvements made to the detector
simulation since (\cite{ourcrab}). The acceptances will be discussed after the analysis cuts have been explained.
Simulations indicate that the trigger rate is dominated by 18~Hz of protons, with 4~Hz of helium
nuclei. Photons from the Crab nebula trigger the detector at about 2\% of the trigger rate.

\begin{table*}
{\centering
\begin{tabular}{ccccccc}
    Source	 &     SP  &    DP      &  SPV   &  All pairs & $<15$ Hz &   On hours  \\ \hline
 Mrk~421         &     61  &    44      &  28    &	133   & 20~\%    &   38.9      \\ 
 Mrk~501	 &     10  &    25      &  --    &       35   & 50~\%    &   10.3      \\ 
 1ES~1426+428     &     --  &    --      &  33    &       33   & 40~\%    &    8.8      \\  
\end{tabular}\par}
\caption{Number of data pairs, after applying stability criteria to the PMT anode currents and to the trigger rates. Single Pointing (SP),
Doubling Pointing (DP), and Single Pointing with Veto (SPV) are explained in the text. The fraction of the data set with a cosmic ray
trigger rate below 15~Hz, which we ascribe to seasonal increases in atmospheric extinction, is also listed.}
\label{datatab}
\end{table*}

\subsection{Optical Throughput}
Previously, the phototube illuminations when viewing stars were lower than predicted by the simulation. We reviewed the ray tracing
programs, and found two factors that dominated the discrepancy. First, the combined heliostat and secondary mirror reflectivities are about
20~\% worse than earlier measurements indicated\,\footnote{We thank Professor Ladislav Rob of Charles University, Prague, and his
colleagues at Compas Ltd., Turnov, Czech Republic, for the recent measurements.}. Second, the simulated heliostat focussing was sharper
than in reality. We compared the image sizes obtained from star ``scans'' used for heliostat alignment with simulated sizes, and modified
the simulation to improve agreement. (In scans, we record PMT anode currents as the heliostats are pointed successively at a grid of points
around the star position: see figure~7 in \cite{repoman}).

\begin{figure}[h]
\resizebox{\hsize}{!}{\includegraphics[width=0.8\textwidth]{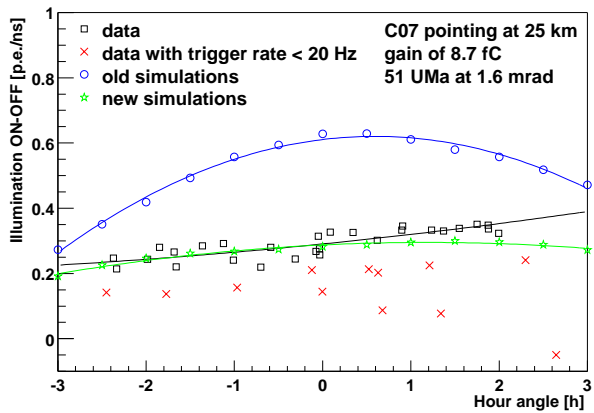}}
\caption{Example of phototube illuminations induced by the star 51 UMa in the field of view of Mrk~421, versus
hour angle, for heliostat C07. Squares \& crosses: For the 40~pairs of On- and Off-source ``double-pointing''
data runs passing current and rate stability criteria, the quantity shown is the difference of the average
current during each run divided by the PMT gain $g$, $(i_\mathrm{On}-i_\mathrm{Off})/g$. For the crosses, the
trigger rates were stable but lower than usual: optical transmission was low, presumably due to clouds or
frost. 
Circles: Original detector simulation. Stars: Simulation results, after
``defocussing'' the heliostats and degrading the mirror reflectivities as described in the text. The curves are
to guide the eye.
\label{EB}} 
\end{figure}

Figure~\ref{EB} illustrates a consequence of the ray tracing modifications (\cite{EBcospar}). 
When tracking Mrk~421, the star 51~UMa (HD~95934, blue
magnitude $M_\mathrm{B}=6.16$) falls within the $\pm 5$~mrad optical field-of-view for heliostats near the center of the array. The
On-source PMT anode currents are thus larger ($i_\mathrm{On} \sim 13~\mathrm{\micro A}$) than the Off-source currents which are due only to
the diffuse night sky light ($i_\mathrm{Off}\sim 10~\mathrm{\micro A}$). Using the PMT gains $g$, the current differences yield the
star-induced illuminations $b = (i_\mathrm{On}-i_\mathrm{Off})/ge$, in photoelectrons per second ($e$ is the electron charge). The figure
shows that measured illuminations vary with the pointing direction, but less than was predicted by the original simulation. The simulations
include the stars in the On and Off fields that contribute at least 5~\% as much as 51~UMa, that is, $M_\mathrm{B} < 9.4$.
We remark that the gamma-ray energy scale predicted by the
Monte Carlo simulation shifts less than figure~\ref{EB} suggests, since an atmospheric shower is a light source of angular extent
comparable to the detector field-of-view, and so the effect of heliostat ``defocussing'' is less violent for showers than it is for the
point-like stars.

A few observations constrain the uncertainty in the optical throughput. Photometry using star scans yields 20~\% channel-to-channel
dispersion compared to the simulated values. A census of damaged heliostat mirrors, and the phototube manufacturer's measurements of the
photocathode responses, combined together account for half of the dispersion. We attribute the rest to errors in the gains, and focussing
anomalies for individual heliostats. However, what matters most is the sums over the 8~heliostats of the trigger groups: the {\em total}
variation over the 5~groups is $\pm 15$~\%, and the rms is less than $10$~\%. We conclude that the overall effect on the gamma-ray energy
scale of the uncertainty in the optical throughput $\delta \epsilon_{opt}$ is less than 10~\%.

Since (\cite{ourcrab}), the electronics simulation has been completely re-written, leading to improved agreement between predicted and
observed signals, in the trigger, and for single channels. The good agreement for the final analysis variable, described below, is
the result of the care taken to model the elements of the complex electronics chain.

\subsection{Atmospheric Transmission}
The amount of Cherenkov light that reaches the ground varies between different sites and seasons depending mainly on the aerosol content
versus altitude (\cite{bernlohr}). The same also holds for atmospheric extinction in stellar photometry (\cite{HL}). The effect on CELESTE
is particularly striking: the $\sim 25$~Hz cosmic ray trigger rates for winter sources ({\em e.g.} Crab and Mrk~421) decrease to under
15~Hz every year in spring, affecting sources like Mrk~501 and 1ES~1426+428. Table~\ref{datatab} lists the fraction of the data sets above
and below 15~Hz.

The CORSIKA Monte Carlo provides a choice of tabulated wavelength and altitude dependent extinction curves. The default contains some
aerosols. The Palomar curve from (\cite{HL}) used for our stellar photometry studies contains fewer aerosols, for an integrated
transmission above Th\'emis at 400~nm 6~\% greater than for the default CORSIKA curve.

An amateur telescope with filters and a CCD camera was used to measure the total integrated atmospheric transmission during some
nights when the trigger rates were above 20~Hz, using the apparent brightness of several stars as a function of their zenith angles
(Bouguer method), at three wavelengths (blue: 390-490~nm, green: 500-560~nm, and red: 610-660~nm). The Cherenkov light recorded by CELESTE
is mainly in the blue range. The CCD results show that the sky at Th\'emis corresponds to the Palomar extinction curve. We made a CORSIKA
extinction table corresponding to the Palomar curve which we use to simulate good running conditions.

To calculate the acceptances for data acquired with trigger rates $<15~\mathrm{Hz}$ we made another CORSIKA extinction table that we call
``dusty''. We increased the aerosol contribution to the Palomar curve so that the integrated transmission above Th\'emis at 400~nm is 60~\%
less. This choice comes from two observations. The first appears in Figure~\ref{EB}, where the phototube illumination due to the bright
star near Mrk~421 in good conditions is halved for data acquired with a low trigger rate. The second is that proton shower simulations
using the nominal and ``dusty'' extinction tables reproduced the experimental nominal and degraded trigger rates, respectively.
Figure~\ref{acceptc}, discussed below, shows the acceptance curves obtained for the two atmospheres, 
used to convert our gamma-ray rates to fluxes.

%For the record, a custom-built LIDAR operated at Th\'emis along with CELESTE ultimately contributed little to our mastery of the
%atmosphere, due to difficulties with the quantitative interpretation of its data.

\begin{figure*}
\centering
\includegraphics[width=0.4\textwidth]{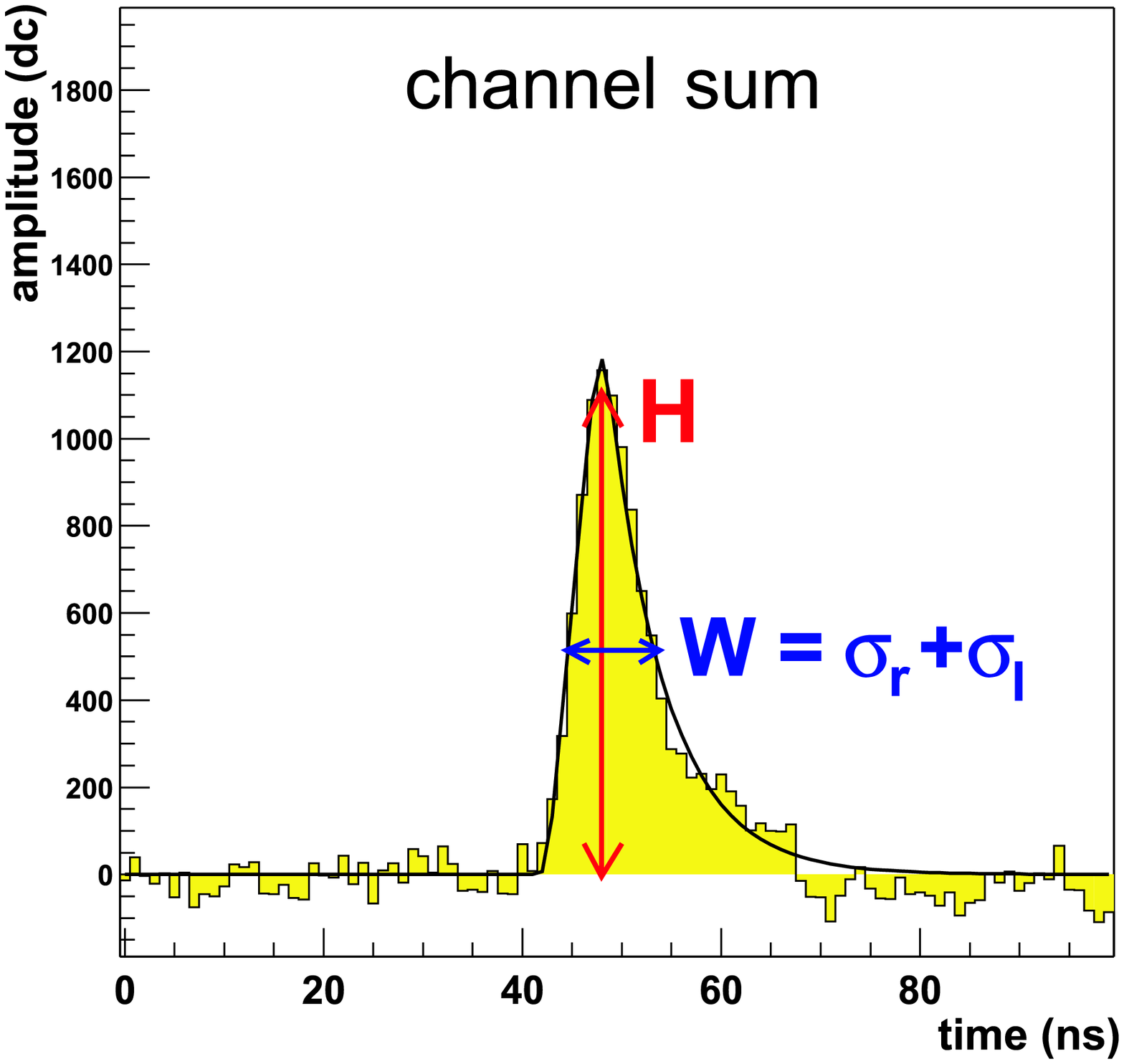}
\includegraphics[width=0.4\textwidth]{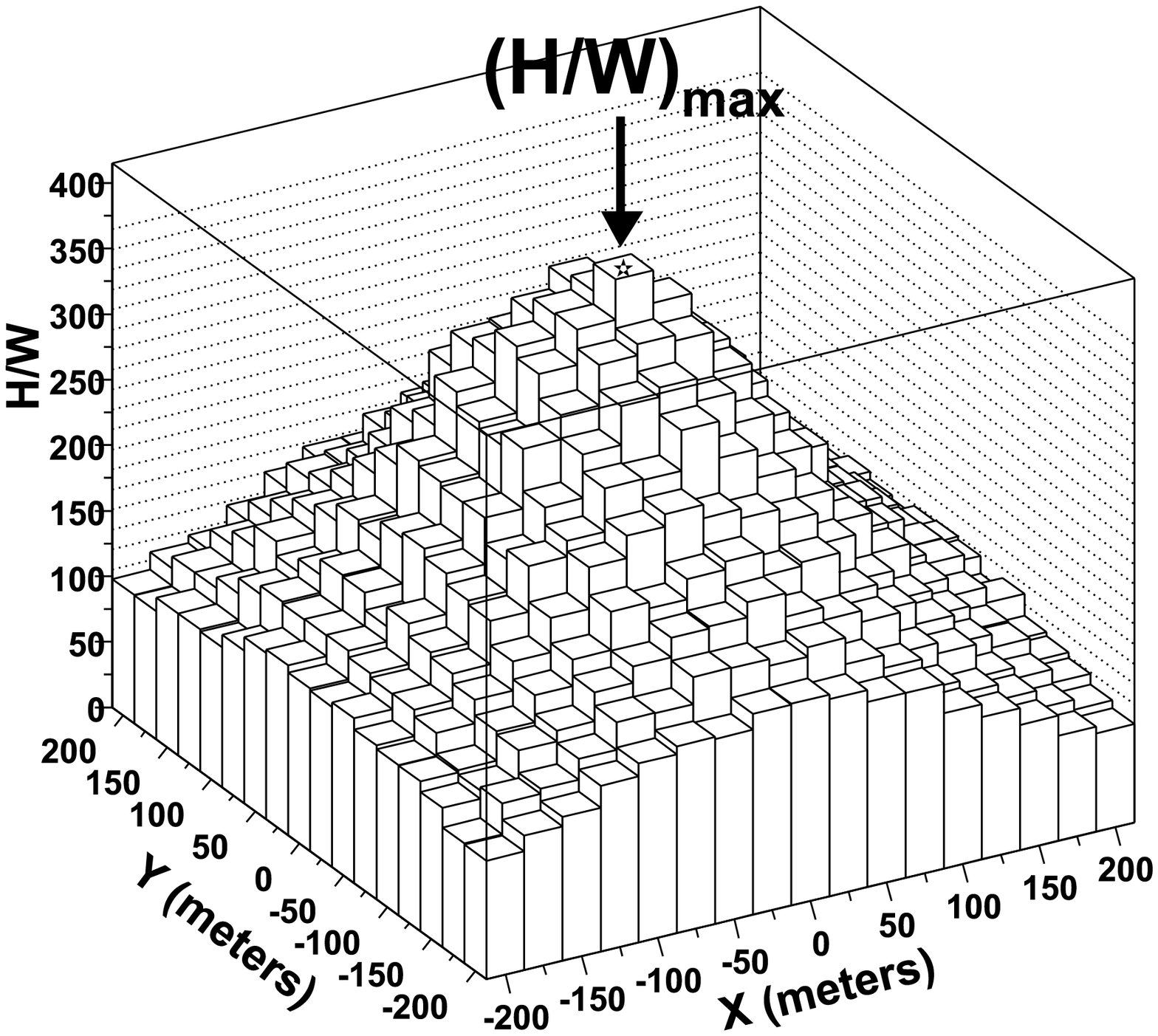}
\caption{LEFT: Summed Cherenkov pulse~-- a 100~ns Flash ADC window, centered at the time an ideal spherical wavefront 11~km directly above
the site would reach the photomultiplier tubes after reflection from the heliostats, is recorded for each channel. In analysis, we
sum the 40 (SP and DP data) or 41 (SPV data) channels repeatedly over a grid of hypothetical shower core positions. 
RIGHT: The $H/W$ values for each position of the grid, for a
single simulated event. The grid position with $(H/W)_\mathrm{max}$ is a good ($\pm 15$~m) estimator of where the gamma-ray would have hit
the ground. The analysis variable $\xi = (H/W)_{200~\mathrm{m}}/(H/W)_\mathrm{max}$ uses $(H/W)_{200~\mathrm{m}}$, which is
the value averaged over a ring 200~meters from the grid center. The vertical axis is in ADC counts per nanosecond.}
\label{HoverL}
\end{figure*}

\section{Data Samples}

Maximum Cherenkov emission for 100~GeV $\gamma$-ray showers occurs around 11~km above the site. For CELESTE's oldest data, the heliostats
were aimed at this height in the direction of the source, to maximize Cherenkov light collection. This is called Single Pointing data (SP).

SP provides the lowest energy threshold, but also decreases the detector's sensitive area. For one season we took data in a configuration
called Double Pointing (DP), where half the heliostats point at 11~km and the rest at 25~km. This increases the area and improves cosmic
ray rejection, while raising the energy threshold only slightly.

Our $\pm 5$~mrad field-of-view matches the apparent size of a gamma shower, to optimize the ratio of Cherenkov light to night sky
background light. Tails of cosmic ray showers are outside the field-of-view, reducing background at the trigger level but weakening offline
rejection. In 2001 we added 13~heliostats to CELESTE, and thereafter pointed 41~heliostats at the 11~km point, and the remaining 12 at a
ring 150~meters around the point. The 12 ``veto'' heliostats can see shower tails outside the central field-of-view, enhancing cosmic ray
rejection (\cite{HMthesis}). We call this Single Pointing with Veto data (SPV). 

Ultimately the performance of the analysis was such that the veto rejection was not used. But the 12 veto heliostats around the edge
of the heliostat field led us to restructure the trigger into $M=6$ groups. The changed trigger topology affects the shape of the recorded
showers, leading to different energy dependent acceptances.

Source observations (``On'') last about 20~minutes and are followed or preceded by an observation at the same declination, offset by
20~minutes in right ascension. ``Off'' data gives a reference for the cosmic ray background~-- the signal is the difference between On and
Off, after analysis cuts. However, changes in sky conditions between On and Off can cause differences having nothing to do with gamma-rays:
we select stable On-Off pairs with a cut on the $\chi^2$ value obtained from a least-squares fit to a constant PMT anode current versus
time, and to the trigger rate versus time, for both runs. We also rejected pairs with very low data rates $(<10~\mathrm{Hz})$, or
anomalously high trigger dead times. The weather at Th\'emis is fickle, and we rejected over half of our data.

Table~\ref{datatab} lists the data sets, recorded during clear, moonless nights from December 1999 through May 2004, 
after data selection.

\section{Data Analysis} 
\subsection{Event reconstruction and background rejection}

Since (\cite{ourcrab}) we have changed our analysis strategy, as described below. The Flash ADC data quality
improved, as did the channel-to-channel timing offsets, which both help wavefront reconstruction. Our procedures to remove the biases induced by night sky background differences at the
trigger level (``software trigger'') and in the FADC data  (``padding'') have been refined. Overall sensitivity has more than doubled.

\begin{figure*}
\includegraphics[width=0.99\textwidth]{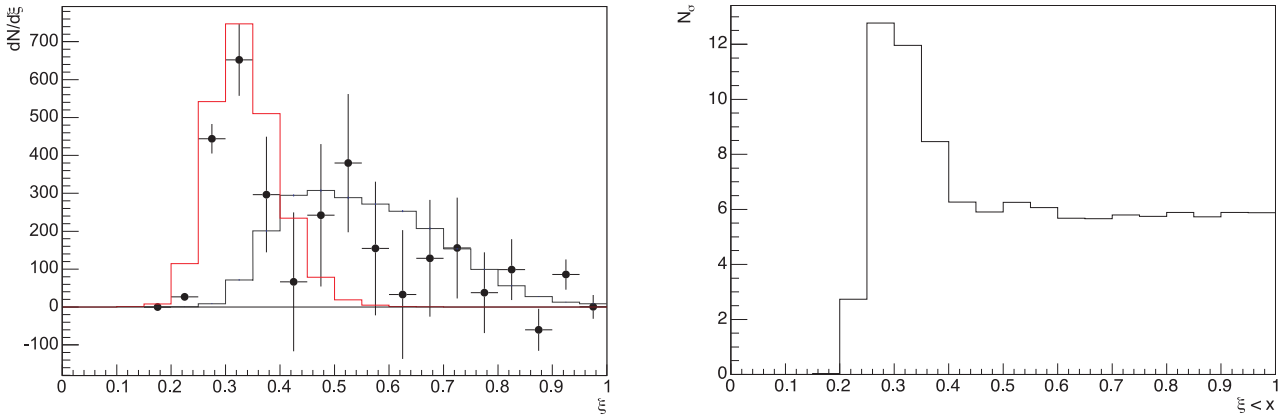}
\caption{LEFT: $\xi$ distributions for Mrk~421 for March 2004 data. Black dots show On-Off
data. The broad histogram is Off data ({\em i.e.} cosmic ray background). The narrow
histogram is simulated gamma-rays. The simulation and Off data are normalized to the
number of On-Off events. 
RIGHT: Statistical significance of the excess as a function of the $\xi$ cut value.}
\label{421xi}
\end{figure*}

Just above the trigger threshold, the Cherenkov signal in one channel is comparable to the fluctuations of the night sky background light.
Summing the channels greatly improves the signal-to-noise ratio, after correcting for the propagation delay to each heliostat, assuming a
spherical Cherenkov wavefront. Lacking knowledge of the shower core position, we scan the plane at 11~km and evaluate the ratio $H/W$ for
each point in a grid, where $H$ and $W$ are the height and the width of the summed signal (see Figure \ref{HoverL}). The grid position
giving the largest value of the ratio $H/W$ is a measure of the shower core position. The wavefront sphericity correlates with how much
$H/W$ decreases 200~m away from the shower core. We measure the relative decrease with a variable called 
$\xi = (H/W)_{200~\mathrm{m}}/(H/W)_\mathrm{max}$,  shown in Figure~\ref{421xi} for an Off
observation, Mrk~421 On-Off data, and for a simulation of a $\gamma$-ray spectrum. The Cherenkov wavefront from $\gamma$-ray induced
showers is, on average, more spherical than for showers initiated by charged cosmic rays and, as expected, $\gamma$-ray showers have
smaller $\xi$ than showers initiated by charged cosmic rays. The method and preliminary results were shown in (\cite{philippe}) and 
detailed in (\cite{HMthesis}).
\\
The $\xi$ distributions change for different heliostat pointings. This article
concerns three blazars with similar declinations, all transiting near zenith at
Th\'emis, so the only changes are the pointing configuration and the hour angle. 
We studied simulations and the strong signals from Mrk~421 and the Crab to
explore the $\xi$ cuts (\cite{EBthesis}). Statistical significance (sensitivity) is optimized by cutting 
a bit above the peak in the gamma ray $\xi$ distribution, {\em e.g.} $\xi < 0.35$ in Figure \ref{421xi}. 
For other pointings, 
the peak lies between $\xi \simeq 0.25$ and $\xi \simeq 0.35$. 
The acceptance depends in part on the efficiency of the $\xi$
cut, obtained from the Monte Carlo. This efficiency is related to the integral of
the $\xi$ distribution, a steep function of $\xi$ near the peak. 
Thus we choose to cut beyond the peak, at $\xi < 0.4$, for all data sets, which increases the statistical uncertainty, but
decreases the acceptance uncertainty $\delta \epsilon_{acc}$ to less than 10\%, and is simpler than having
an array of cut values for the various pointings.

\subsection{Acceptances and Performance}
 
\begin{figure*}
   \includegraphics[width=0.95\textwidth]{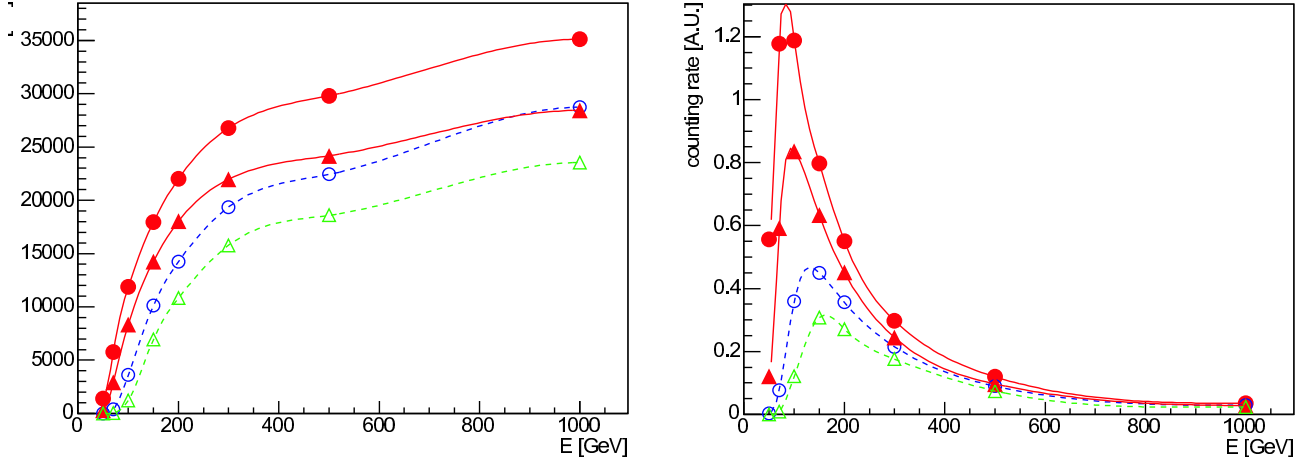}
   \vspace*{-10.5cm}
\caption{LEFT: Energy dependent gamma-ray collection area $A(E)$, at the trigger level (round markers), 
and after analysis cuts (triangles), for sources culminating near zenith, 1~hour after transit, for 11~km single pointing (SP). 
The curves are splines. The two solid curves used the
low-aerosol atmosphere while the two dashed curves used the ``dusty'' atmosphere (see text). 
The analysis trigger group multiplicity cut for these plots was $M=4$ of the $N=5$ groups. 
RIGHT: $A(E)$ multiplied by a typical $E^{-2}$ flux spectrum, to give the relative counting rate $A(E)/E^2$. 
Integral fluxes are quoted at the energy corresponding to the maximum rate, while the energy range of
50 to 350 GeV corresponds to 20\% of the maximum rate.
\label{acceptc}}
  \end{figure*}

For a gamma-ray source with differential spectrum $\phi(E)$ the 
number $N$ of detected gammas will be $$ N = T\int_0^\infty A(E) \phi (E) \mathrm{d}E$$ where $T$ is 
the On-source time and $A(E)$ is the energy dependent gamma-ray collection area obtained from the Monte Carlo calculations. 
We have calculated $A(E)$ for each heliostat configuration, and for hour angles of $(0, 1.0, 1.5, 2.0)$~hours, 
for each of the two atmospheres described above. Figure~\ref{acceptc} shows $A(E)$ at the trigger level, and after cuts, 
for the direction of the blazar Mrk~421 ($\delta=38\degr$), one hour after transit. 
(The few degree discrepancy with the other two blazar's declinations induces a negligible error). 
For the blazar spectra we assume a power law, $\phi(E)=kE^{-\alpha}$.
CELESTE's energy range is near the SED high energy peak for 
the three blazars, where $\alpha \equiv 2$. We vary $\alpha$ between $1.8$ and $2.2$ to estimate the uncertainty 
due to this choice. The energy reconstruction method 
in (\cite{fred}) was used to search for a signal in energy bands in (\cite{M31}) but was not applied here.

The combined uncertainties 
$(\delta\epsilon_{MC} \otimes \delta\epsilon_{opt})<$ 20\% affect the energy scale.
Furthermore, along with $\delta\epsilon_{acc}$ they lead to a flux error.
A remaining uncertainty comes from the varying atmospheric transmission. 
We make a first-order correction by grouping the data into subsets with trigger rates 
of $\leq 15$ and $>15$~Hz (see Table~\ref{datatab}) for use with the nominal or ``dusty'' acceptances, 
$A(E)$, as in Figure \ref{acceptc}. The maximum counting rate after cuts is nearly halved
with the ``dusty'' atmosphere. The observed extrema of the trigger rates are 12 and 24~Hz, 
but the distribution clusters around 14 Hz and 22 Hz. 
We conclude that the overall systematic flux uncertainty is $<25$\%.
%
%$\langle A\rangle = \displaystyle\frac{\int_0^\infty A(E) \phi (E) \mathrm{d}E}{\int_0^\infty \phi (E) \mathrm{d}E} $ 
%The dependence of $\langle A\rangle$ on the spectral index $\alpha$ further biases the flux, 
%by less than 20~\% for $\alpha=2 \pm 0.2$.

To summarize CELESTE's performance: for one hour of On-source Crab data near
transit, analysis using $\xi < 0.4$ gives $3.7 \pm 0.2$~gamma/min over the whole data
set. The significance is $3.3~\sigma$ for SP and
DP data, and $5~\sigma$ for SPV. The higher sensitivity for the more recent data
is due mainly to having six trigger groups instead of five, at the cost of 
a somewhat higher minimum energy: Table \ref{results421tab} shows that the SP raw data rate is
15\% higher than for SPV, but that analysis cuts reject more background for SPV data (90\%) than
for SP data (83\%). The higher sensitivity also stems
in part from improvements in FADC quality, timing adjustments, and gain
settings over the years. Optimized cuts increase the significances by 20~\%, and more for
Mrk~421 (\cite{EBthesis}) but are not used in this paper.  

\begin{figure*}
\centering
\includegraphics[width=0.95\textwidth]{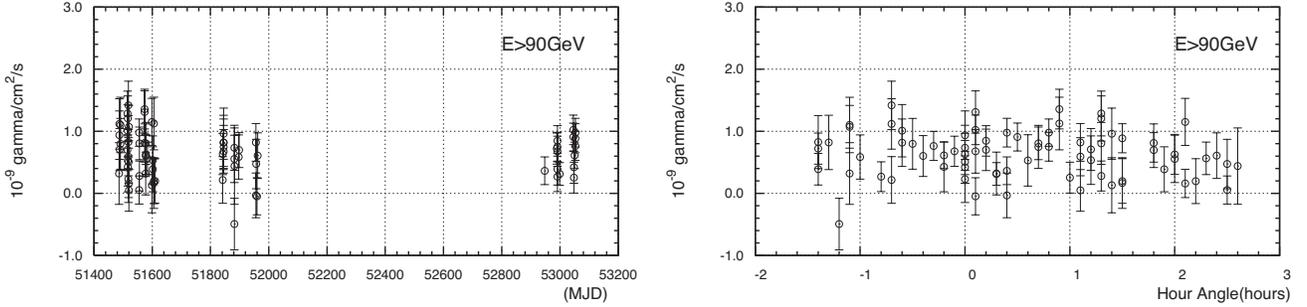}
\caption{LEFT: Crab fluxes for single data runs over the 5 years that CELESTE ran. 
SP pointing was mainly used during the first year, DP for the 2nd year, and SPV for the last. 
The integral flux above 90 GeV is  
$0.62 \pm 0.04 \times 10^{-9}$ photons/cm$^{2}$/s averaged over the three good Crab seasons. 
RIGHT: Same, as a function of the hour angle of the data runs. After acceptance corrections
the observed flux is steady.}
\label{crab}
\end{figure*}

Folding our energy dependent acceptance $A(E)$ (similar to Figure~\ref{acceptc}) with the Crab inverse
Compton spectrum parametrized in Table~6 of (\cite{hegraCrab}) yields a maximum at
90~GeV. The rate falls to 20~\% of the maximum below 50~GeV and above 350~GeV. 
Figure \ref{crab} shows the acceptance-corrected pair-by-pair Crab fluxes, as a function of date and hour angle. 
%The RMS of the normalized pair-by-pair residuals about the average rate is 1.2, where a value of 1
%indicates that the light curves are free of systematic biases. 
The stability is adequate to search for blazar variability. 
%The average integral flux above 90 GeV is $0.62 \pm 0.04 \times10^{-9}$~cm$^{-2}$\,s$^{-1}$.
The power flux at 90 GeV is  $\nu F_\nu = 0.89 \pm 0.05 \times 10^{-10}$~erg cm$^{-2}$\,s$^{-1}$, 
higher than our previous result in (\cite{ourcrab}) but within the uncertainties, and 
in agreement with the spectrum in (\cite{hegraCrab}).

\section{Results and Discussion}

\begin{figure*}
\hspace*{-2.5cm}
\centering\includegraphics[width=0.7\textwidth, angle=270]{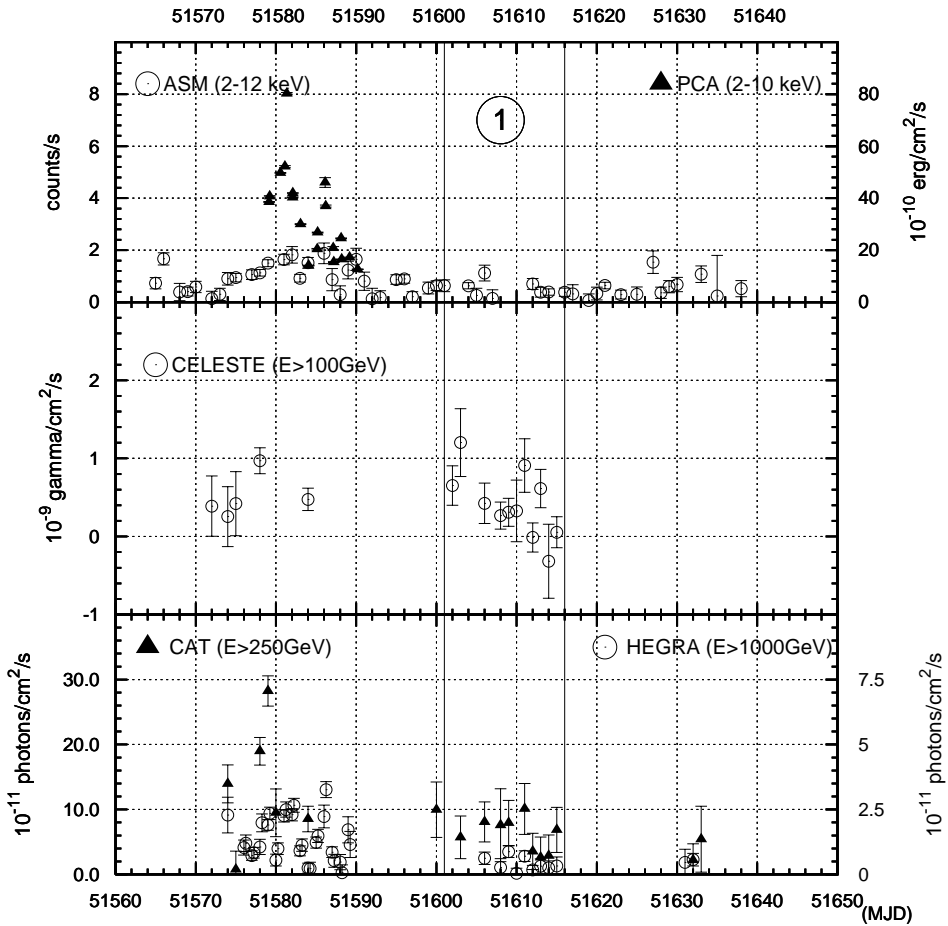}
\caption{Mrk~421 daily averages in early 2000. 
The CELESTE data runs from MJD 51572 to 51613 (January 29 to March 10), while epoch
(1) is defined in Table \ref{421FluxTab} and discussed in the text.
CAT data is from (\cite{cat421}, left-hand scale) 
while the HEGRA points are from (\cite{feb2000}, right-hand scale).
RXTE ASM (\cite{ASM}, left-hand scale) and PCA (\cite{feb2000}, right-hand scale)
results are also shown. 
}
\label{421lc2000}
\end{figure*}

\begin{figure*}
\hspace*{-2.5cm}
\centering\includegraphics[width=0.7\textwidth, angle=270]{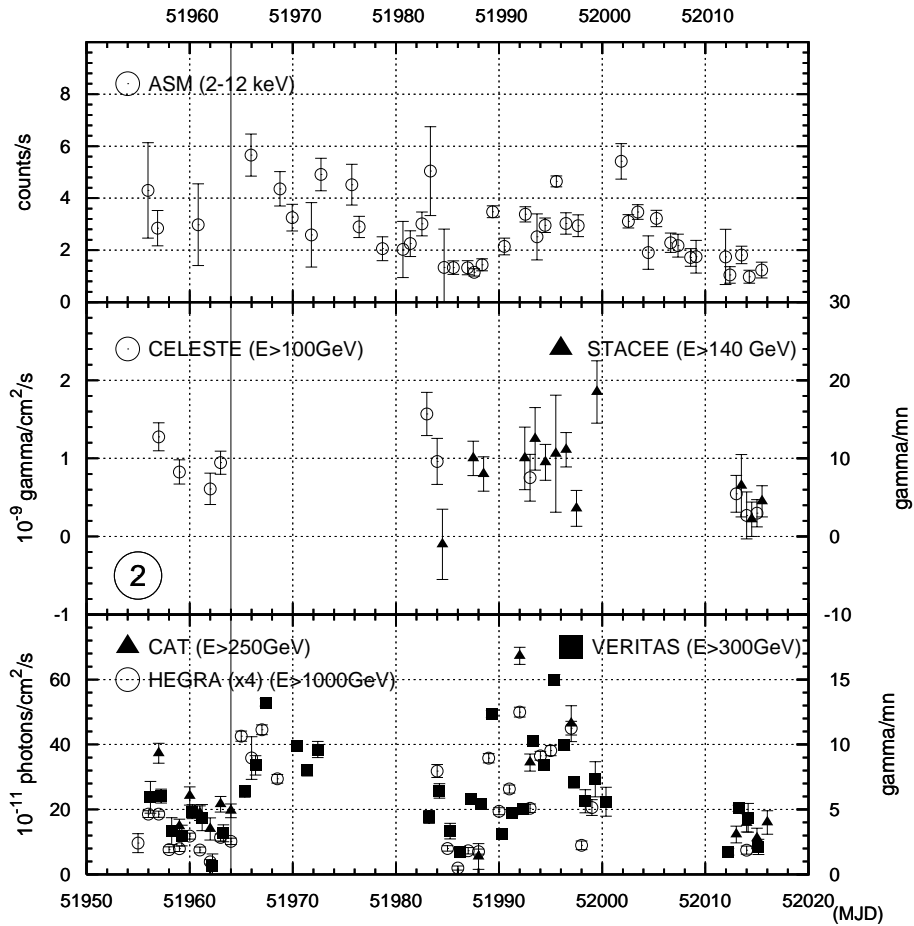}
\caption{Mrk~421 daily averages in 2001. Epoch
(2) is defined in Table \ref{421FluxTab} and discussed in the text. 
CELESTE data runs from MJD 51957 to 52015 (February 17 to April 16). 
STACEE data is from (\cite{stacee421}), HEGRA data is from (\cite {hegra421a}), and
the VERITAS data is from (\cite{jamie}). The other data points are as in the preceding figure.
Note the different left- and right-hand axes.}
\label{421lc2001}
\end{figure*}

\begin{figure*}
\hspace*{-2.5cm}
\centering\includegraphics[width=0.7\textwidth, angle=270]{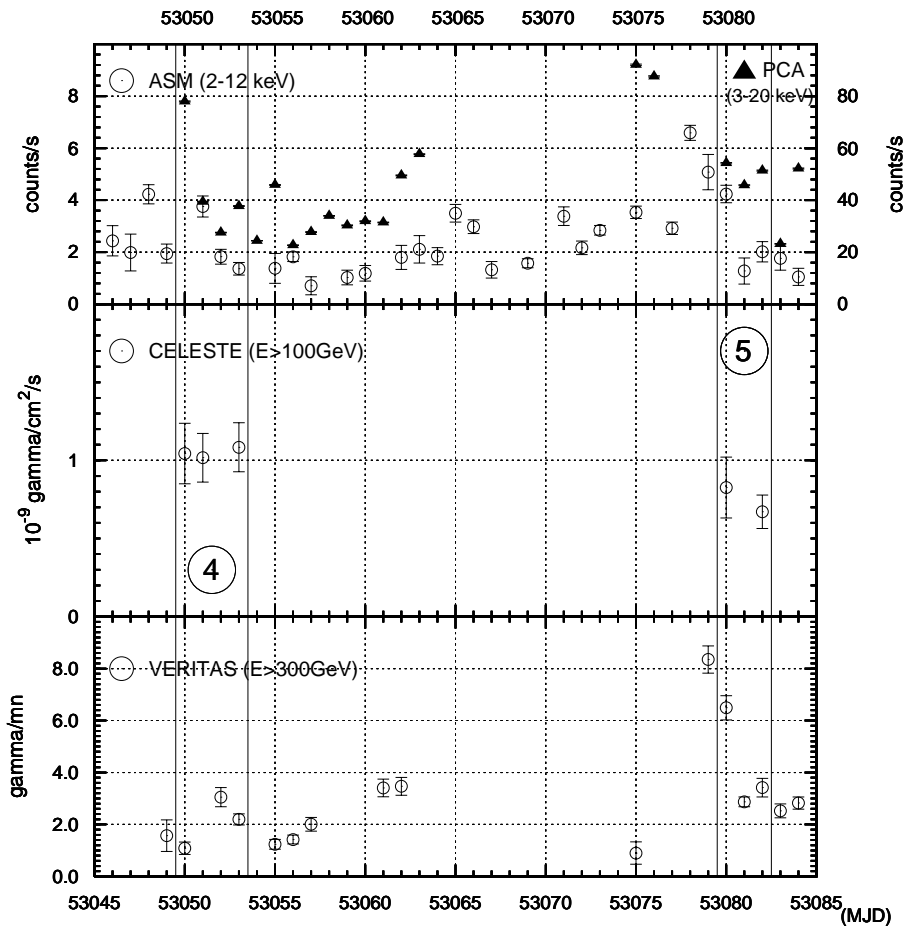}
\caption{Mrk~421 daily averages in February and March 2004. 
Epochs (4) and (5) are defined in Table 3 and discussed in the text.
The RXTE PCA and VERITAS data
are taken from (\cite{blaze}), while the ASM data comes from (\cite{ASM}).}
   \label{421lc2004}
\end{figure*}

Tables~\ref{results421tab}, \ref{results501tab}, and \ref{results1426tab} summarize the data 
analysis results for the three blazars Mrk~421, Mrk~501, and 1ES~1426+428, respectively.
Figures~\ref{421lc2000},~\ref{421lc2001}, and \ref{421lc2004}, 
show the acceptance-corrected light curves for the 
blazar Mrk~421 during the years 2000, 2001, and 2004 seasons.
Figures \ref{501lcDAS} and \ref{1426lc} show the light curves
for Mrk~501 and 1ES~1426+428, respectively. The curves assume power laws with spectral index $\alpha = 2$. 
There is an often strong signal from Mrk~421, and no signal for 1ES~1426+428. 
The excess in the Mrk~501 data is discussed below.

\begin{table*}
\begin{center}
\resizebox{\hsize}{!}{
\begin{tabular}{llllllll}
 &  & \multicolumn{3}{c}{Number of events} & Significance & Signal-to-noise & Excess\\
Data set & Cut & $N_\mathrm{\scriptsize{On}}$ & 
$N_\mathrm{\scriptsize{Off}}$ & 
$N_\mathrm{\scriptsize{On}}-N_\mathrm{\scriptsize{Off}}$ 
&  $N_\sigma$ & 
$\frac{N_\mathrm{\scriptsize{On}}-N_\mathrm{\scriptsize{Off}}}{N_\mathrm{\scriptsize{Off}}}~\mathrm{[\%]}$
& $\mathrm{[\gamma/\mathrm{min}]}$\\
\hline 
SP ($17.8~\mathrm{h}$) & Raw Data & $1\,575\,396$ & $1\,557\,192$ & $18\,204$ &  & & \\
& Software trigger & $1\,056\,705$ & $1\,044\,811$ & $11\,894$ & 
$7.1$ & $1.1$ & $11.1$\\
& All cuts & $276\,057$ & $270\,113$ & $5\,944$ & $7.1$ & $2.2$ & $5.6$\\
\hline
DP ($12.9~\mathrm{h}$) & Raw Data & $983\,550$ & $961\,257$ & $22\,293$ &  &  & \\
& Software trigger & $532\,883$ & $523\,661$ & $9\,222$ & $8.1$ & 
$1.8$ & $11.9$\\
& All cuts & $113\,507$ & $106\,904$ & $6\,603$ & $12.8$ & $6.2$ & $8.5$\\
\hline
SPV ($8.2~\mathrm{h}$) & Raw Data & $621\,243$ & $604\,926$ & $16\,317$ & &  &\\
& Software trigger & $410\,429$ & $401\,791$ & $8\,638$ & $8.5$ & 
$2.1$ & $18.3$\\
& All cuts & $62\,914$ & $57\,551$ & $5\,363$ & $13.9$ & $9.3$ & $11.8$\\
\hline \hline
All data ($38.9~\mathrm{h}$) & Raw data & $3\,180\,189$ & $3\,123\,375$ & $56\,814$ & &  &\\
& Software trigger & $2\,000\,017$ & $1\,970\,264$ & $29\,754$ & 
$13.1$ & $1.5$ & $12.7$\\
& All cuts & $452\,478$ & $434\,568$ & $17\,910$ & $16.9$ & $4.1$ & $7.7$\\
\end{tabular}
}
\caption{Number of events before and after cuts, for the 
Mrk~421 data sets.
$N_\sigma$ is calculated after correcting for the $\sim 80\%$ data acquisition
efficiency.
}
\label{results421tab}
\end{center}
\end{table*}

\begin{table*}
\begin{center}
\resizebox{\hsize}{!}{
\begin{tabular}{lllllll}
Epoch & ASM (a) & PCA & CAT (d) & HEGRA (e)  & VERITAS (f) & CELESTE (g)\\
\hline 
2000 Feb 4 & $1.25\pm0.10$  & & $189.7\pm21.2$ & $11.24\pm3.41$ & & $0.97\pm0.17$\\
(51577.99 to 51578.18) & & & & & &\\
2000 Feb 10 & $1.51\pm0.21$  & $1.41\pm0.02$ (b) & $85.3\pm19.7$ & $2.43\pm1.93$ & & $0.48\pm0.14$\\
(51583.99 to 51584.18) & & & & & &\\
(1) 2000 Feb 27 to March 12 & $0.39\pm0.06$ & & $48.3\pm10.9$ &  $3.9\pm0.8$ & & $0.29\pm0.07$\\
(51601.96 to 51615.09) & & & & & &\\ \hline
(2) 2001 Feb (4 nights)    & $2.98\pm1.57$  & & $208.7\pm10.5$ &  $22.5\pm0.8$ & $4.35\pm0.29$ & $0.93\pm0.09$\\
(51956.98 to 51963.11) & & & & & &\\
2001 April 14 to 17 & $1.26\pm0.17$ & & $131.7\pm16.3$ &  $18.7\pm3.1$ & $4.65\pm0.22$ & $0.34\pm0.12$\\
(52012.90 to 52015.97) & & & & & &\\  \hline
(3) 2003 Feb to April & $0.70\pm0.03$ & $17.3\pm0.02$ (c) & & & $0.94\pm0.04$ & $0.60\pm0.12$\\
(52667.12 to 52762.94) & & & & & &\\  \hline
(4) 2004 Feb 15 to 18 & $2.47\pm0.23$ & $39.3\pm0.1$ (c) & & & $1.66\pm0.20$ & $1.04\pm0.10$\\
(53050.05 to 53053.08) & & & & & &\\
(5) 2004 March 16 to 18 & $1.74\pm0.31$ & $48.4\pm0.2$ (c) & & & $3.37\pm0.17$ & $0.71\pm0.09$\\
(53079.91 to 53082.02) & & & & & &\\
\end{tabular}
}
\caption{Multiwavelength fluxes from Mrk~421 coincident with CELESTE data. The epoch numbers (1) through (5) are
referred to in the text and figures.
(a) ASM counts per second (\cite{ASM}). 
RXTE PCA 2-10 keV fluxes are in (b) ~$10^{-10}~\mathrm{erg\,cm^{-2}\,s^{-1}}$, 3-20~keV (\cite{feb2000}), and in
(c) counts per second (\cite{blaze}).
(d) CAT fluxes are in $10^{-12}~\mathrm{photons\,cm^{-2}\,s^{-1}},\,E>250~\mathrm{GeV}$ (\cite{cat421}, 2001b). 
(e) HEGRA fluxes are in $10^{-12}~\mathrm{photons\,cm^{-2}\,s^{-1}},\,E>1000~\mathrm{GeV}$ 
(\cite{feb2000} for the 2000 data, \cite{hegra421b} for 2001). 
(f) VERITAS fluxes above 300 GeV are in gamma/min, where the Crab gave $2.40$ (2.93) gamma/min for 02/03 (03/04)
 (\cite{jamie,blaze}). 
(g) CELESTE fluxes are in $10^{-9}~\mathrm{photons\,cm^{-2}\,s^{-1}},\,E>100~\mathrm{GeV}$.
\label{421FluxTab}}
\end{center}
\end{table*}

\begin{figure*}
%\resizebox{\hsize}{!}{\includegraphics[width=0.7\textwidth]{./Mrk421SpectreBlazejowski.eps}}
\includegraphics[width=0.7\textwidth]{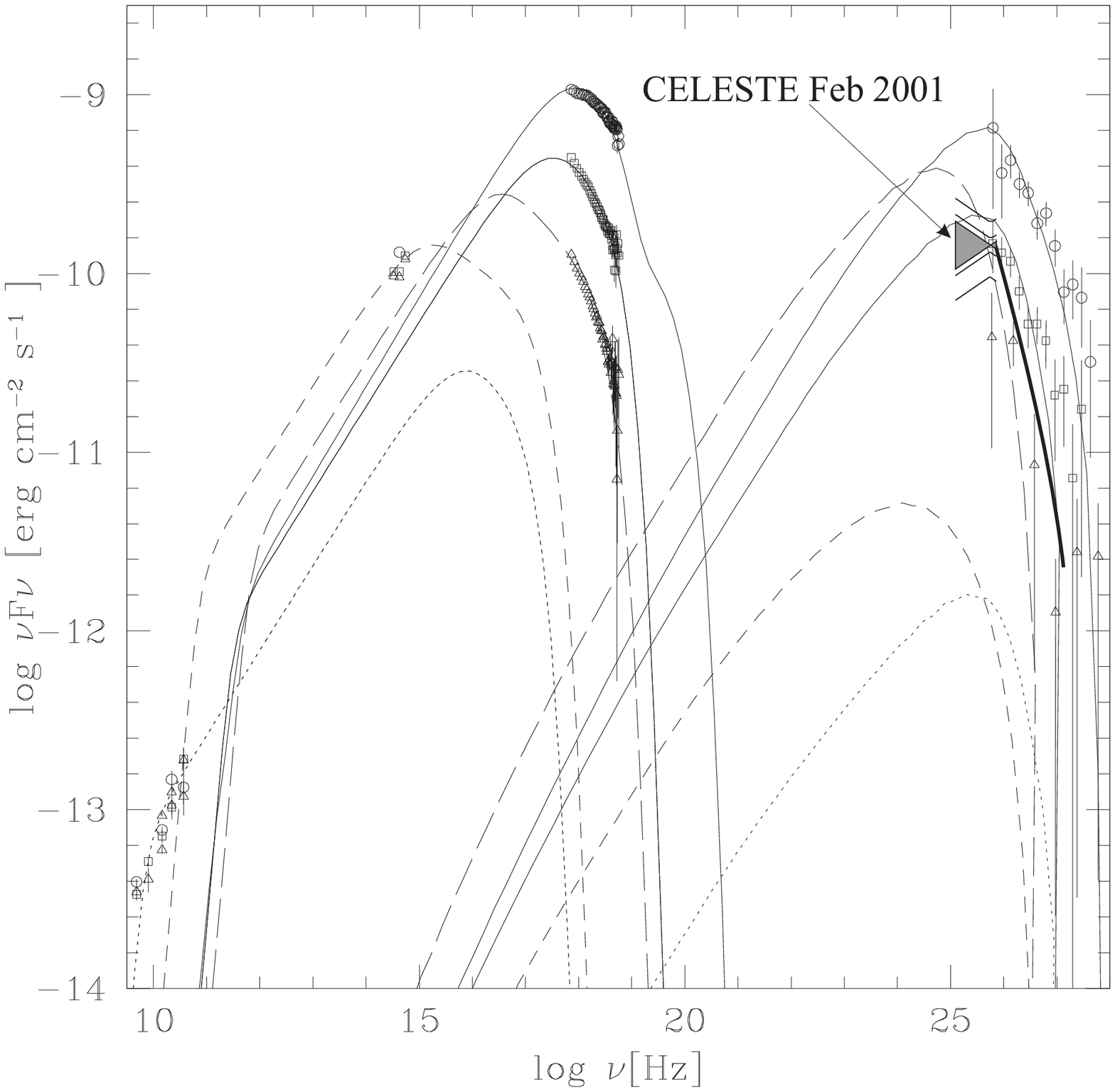}
%\vspace*{-5cm}
\caption{CELESTE average flux
measurement (gray bowtie) for Mrk~421 for 2001 February (MJD 51956.98 to 51963.11,
Epoch (1) in Table 3 and text),
superimposed on the spectral energy distribution taken from the multiwavelength
study by (\cite{blaze}). The envelope of the bowtie corresponds to the assumptions
of $\alpha=1.8$ and $2.2$ differential spectral indices. The inner set of lines above
and below the bowtie are the statistical errors, while the outer set has in addition
a 25\% systematic uncertainty added in quadrature.
The ASM X-ray activity level in February 2001 corresponds 
to the ``medium'' activity state (the lower of the solid-curved high energy peaks) and
thus to the VERITAS TeV measurements represented by the squares.
The CAT spectrum in (\cite{cat421}) for the year 2000 (thick black line) gives the same integral
flux as for the nights in February 2001.
\label{spec421}} \end{figure*}

\subsection{Mrk~421}

Understanding blazars requires observations at different wavelengths, preferably simultaneous given their
variability. Happily, much of our data coincides with previously published results.  
Table~\ref{421FluxTab} lists the CELESTE, TeV, and X-ray fluxes for the epochs with CELESTE data. 
CELESTE's richest data taking period was
March 2000, with 11 of 14 consecutive nights (epoch (1), from 2000 February~27 to
March~12, MJD 51601.96 to 51615.09). Figure~\ref{421lc2000} shows the daily averaged data during this period.
CAT (\cite{cat421}) and HEGRA (\cite{hegra421a}) have data for most of the same
nights, and the four experiments with data at this epoch are all at a low level. 
The CELESTE flux averaged over the 2003 season, epoch (3), from February through April (MJD
52667.12 to 52762.94) was at an intermediate level, as were the other experiments. Overall,
the flux variations observed with CELESTE track the other measurements.

Figure~\ref{spec421} shows the SED for Mrk~421 taken from (\cite{blaze}) to which 
we have added our result for epoch (2) when TeV data from Cherenkov imagers are available,
4 nights in February 2001 (MJD 51956.98 to 51963.11).
ASM indicates that the blazar was at the low end of the ``medium'' X-ray state as defined in (\cite{blaze}). 
We have plotted the CAT spectrum for the year 2000 (\cite{cat421}), which has the same
integral flux as measured on those four nights.

We now look more closely at the light curves. 
The detailed multi-wavelength study reported in (\cite{blaze}) makes February-March 2004 particularly interesting
(epoch 4). 
CELESTE measured a steady high state of $1.04 \pm 0.1\times 10^{-9}$~cm$^{-2}$s$^{-1}$ for the three nights 
of 2004 February~15 to 18 (MJD 53050.05 to 53053.08,
see Table \ref{421FluxTab} and Figure~\ref{421lc2004}). VERITAS shows
a steady low state, in seeming contrast to CELESTE. But Figure~1 of (\cite{blaze}) shows an increasing optical flux on
those dates, while PCA and ASM on RXTE show a decreasing X-ray state. This could be a shift of the low energy peak of
the SED to lower energy. 
%(Figure~1 of (\cite{blaze}) also shows that the radio flux is falling slowly, consistent with
%the end of a flare as seen far from the low-energy peak. In Figure~\ref{spec421} the dotted line through the radio data
%comes from assuming a different acceleration zone than for the medium state X-ray and GeV-TeV curve, as does the dashed
%line through the optical data.) 
If the high energy SED peak also shifts during this epoch it could explain that CELESTE
sees a high rate when VERITAS does not: VERITAS is above the SED peak while the CELESTE range straddles it. 

Similarly, and also in Figure \ref{421lc2004}, PCA and ASM show a declining X-ray state followed by a small rebound for
2004 March~16 to 18 (epoch 5). For two nights of data (MJD 53079.91 to 53082.02) CELESTE has an average state of 
$1.4 \pm 0.2\times 10^{-9}$~cm$^{-2}$s$^{-1}$, but here VERITAS sees the decline of the flare before returning to the low state. 
CELESTE's stability during VERITAS' changes is again consistent with Figure \ref{spec421}, that is, with an SED peak at 100~GeV.
Recent spectral work on Mrk~421 by STACEE also seems to indicate that our 50 to 350~GeV energy range matches a high energy SED peak
(\cite{CarsonThesis}). 

\begin{table*}
\begin{center}
%\resizebox{\hsize}{!}{
\begin{tabular}{lllllll}
 &  & \multicolumn{3}{c}{Number of events} & Significance  & Excess\\
Data set & Cut & $N_\mathrm{\scriptsize{On}}$ & $N_\mathrm{\scriptsize{Off}}$ & $N_\mathrm{\scriptsize{On}}-N_\mathrm{\scriptsize{Off}}$ & $N_\sigma$ & $\mathrm{(\mathrm{min}^{-1})}$\\
\hline 
Year 2000 ($8.3~\mathrm{h}$) & Raw data& $616\,194$ & $613\,337$ & $2\,857$ & &\\
& Software trigger & $362\,863$ & $359\,914$ & $2\,949$ & $2.9$  & $5.9$\\
& All cuts & $73\,034$ & $71\,518$ & $1\,516$ & $3.4$ &  $3.1$\\  \hline
Year 2001 ($2.0~\mathrm{h}$) & Raw data& $130\,820$ & $131\,584$ & $-764$ & &\\
& Software trigger & $67\,040$ & $67\,410$ & $-370$ & $-0.9$ & $-3.1$\\
& All cuts & $13\,957$ & $14\,059$ & $-102$ & $-0.6$  & $-0.85$\\
 \hline
All data ($10.3~\mathrm{h}$) & Raw data& $747\,014$ & $744\,921$ & $2\,093$ & &\\
& Software trigger & $429\,903$ & $427\,323$ & $2\,580$ & $2.4$  & $4.2$\\
& All cuts & $86\,991$ & $85\,578$ & $1\,413$ & $2.9$ & $2.3$\\
\end{tabular}
%}
\caption[]{Number of events before and after cuts for the Mrk~501 data sets. 
$N_\sigma$ is calculated after correcting for the $\sim 80\%$ data acquisition
efficiency.
}
\label{results501tab}
\end{center}
\end{table*}

\subsection{Mrk~501}

\begin{figure*}[t]
\centering
\includegraphics[width=0.99\textwidth]{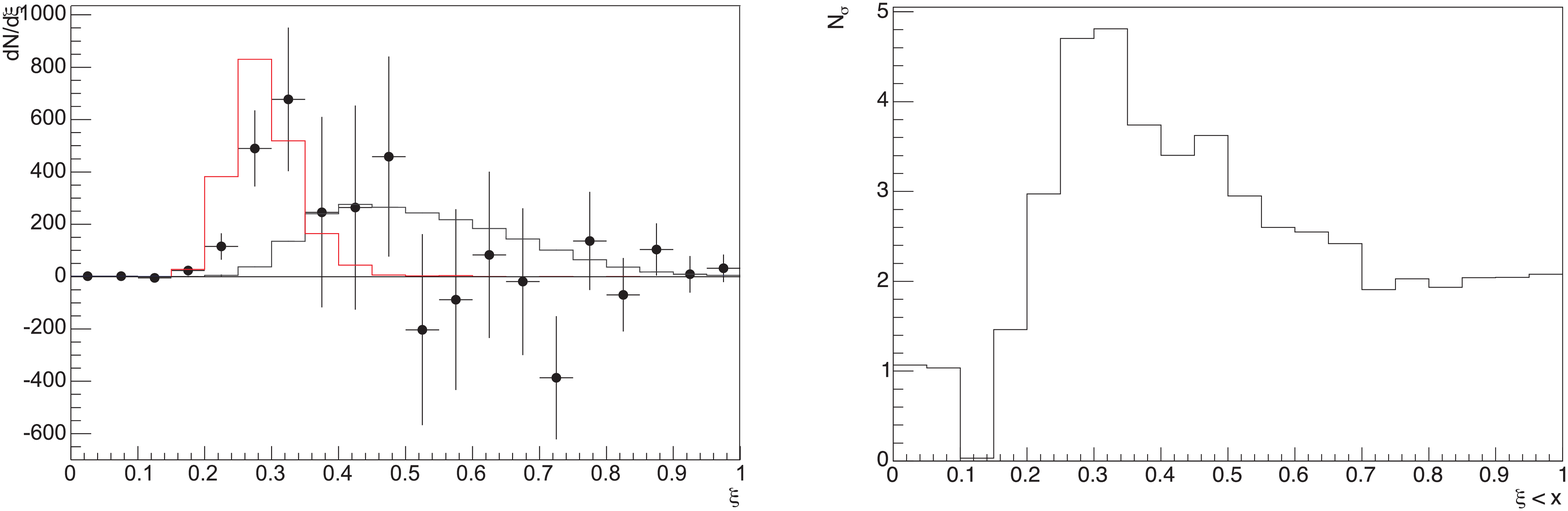}
\caption{Left: $\xi$ distributions for Mrk~501 for the year 2000 data. Black dots show On-Off data. 
The broad histogram is Off data ({\em i.e.} cosmic ray background). The narrow histogram is simulated gamma-rays.
The simulation and Off data sets are normalized to the number of On-Off events.
Right: Statistical significance of the excess as a function of the $\xi$ cut value.}
   \label{501xi}
\end{figure*}

\begin{figure*}
\hspace*{-2.5cm}
\centering\includegraphics[width=0.7\textwidth, angle=270]{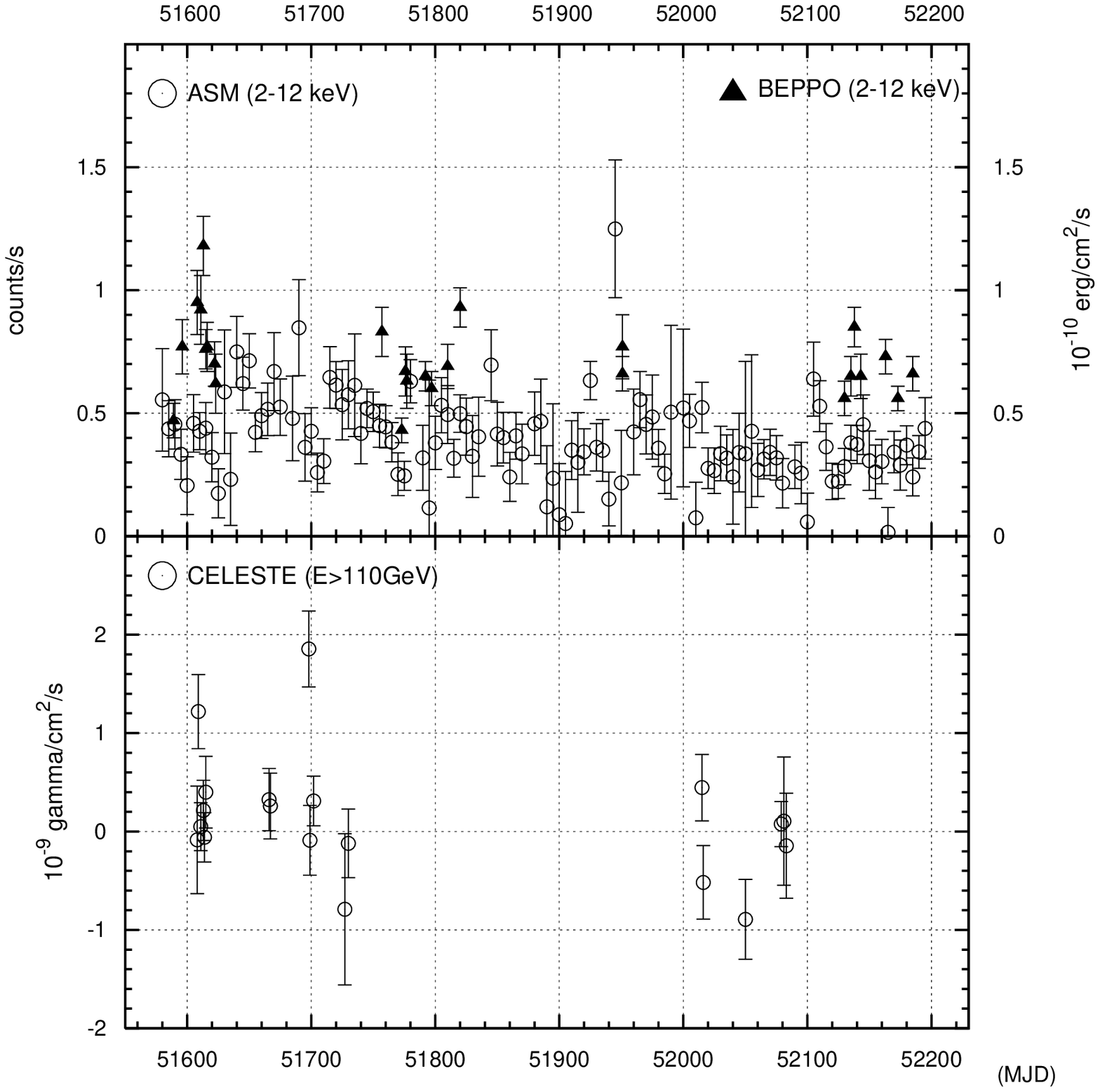}
\caption{Gamma- and X-ray light curves for Mrk~501. $\mathrm{MJD}$ $51600$ and $52000$ correspond to 
2000 February 26 and 2001 April 4.
TOP: ASM 5-day running averages of the 2 to 12~keV X-ray flux, in cts/s (\cite{ASM}), and BeppoSax 2 to
12~keV fluxes in $10^{-10}~\mathrm{erg\,cm^{-2}\,s^{-1}}$, from (\cite{beppo}). BOTTOM: CELESTE daily
flux averages.  }
\label{501lcDAS}
\end{figure*}

The statistical significance of the excess for all the Mrk~501 data is weak ($2.9~\sigma$), as
shown in Table~\ref{results501tab}. The same $\xi<0.4$ cut as for Mrk~421 was applied.
The light curve in Figure~\ref{501lcDAS} shows that the
excess comes mostly from the year 2000, with $3.4~\sigma$ for those four months.
Figure~\ref{501xi} shows that the excess in 2000 ressembles a gamma-ray signal. However,
changing the $\xi$ cut values to those that optimize the statistical significance of the
Mrk~421 signal do not enhance this excess, suggesting that taking the excess to be a gamma-ray
signal is delicate.

\begin{figure*}
\includegraphics[width=0.5\textwidth,angle=270]{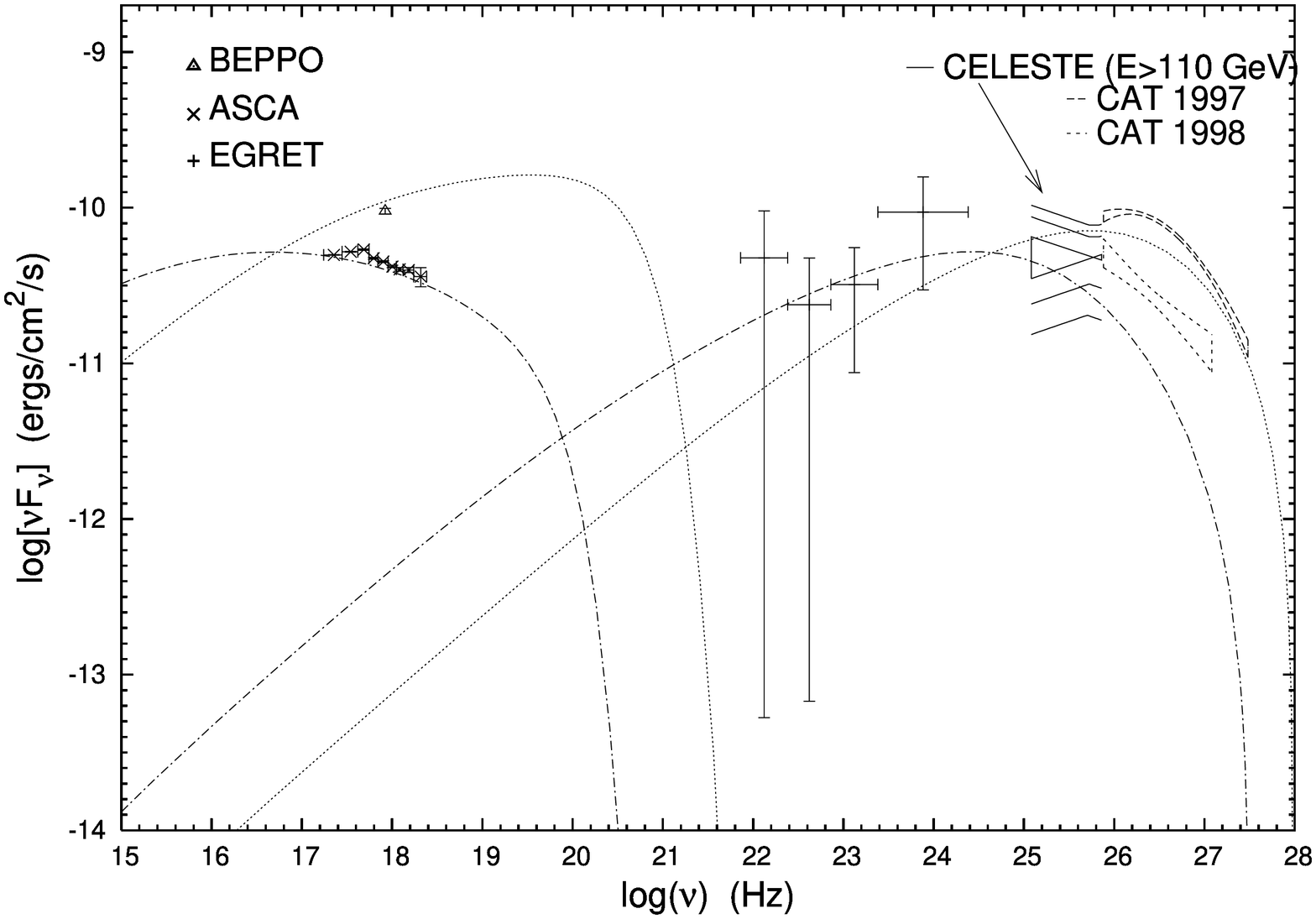}
%\resizebox{\hsize}{!}{\includegraphics[width=0.7\textwidth,angle=270]{./SED_501.eps}}
\caption{CELESTE flux measurement for Mrk501 from the year 2000 (bowtie), bounded
by the assumptions of $\alpha=1.8$ and $2.2$ differential spectral indices. 
The inner set of lines, above and below the bowtie, correspond to the statistical 
uncertainty, while the outer set have in addition the 25\% systematic uncertainty added in quadrature.
The Beppo X-ray data is nearly simultaneous (\cite{beppo}). 
The EGRET data and the ASCA X-ray measurements are 
simultaneous with each other, and correspond to the SSC model curves (\cite{egret501}). 
The TeV spectra are from the CAT Cherenkov imager (\cite {cat501bis}). }
\label{spec501}
 \end{figure*}

CELESTE and the BeppoSAX X-ray satellite have data in common for four nights, from 2000
March~5 to 12, when Beppo measured between $(0.76 \pm 0.08)$ and $(1.18 \pm 0.12) \times
10^{-10}~\mathrm{erg\,cm^{-2}\,s^{-1}}$ (\cite{beppo}). The ASM rate during this period is
a roughly constant $0.4 \pm 0.1$.
(ASM rates are unreliable at such low fluxes.)
%, nevertheless Figure \ref{501lcDAS} shows that they track the Beppo measurements.)
Averaging the ASM count rates for the nights with CELESTE data gives $0.44 \pm 0.06$ and $0.37
\pm 0.09$ for 2000 and 2001, respectively. The dispersion of the points is $0.5$~cts/s for
both seasons. Averaging ASM over longer periods gives essentially the same result, so that no
significant difference in X-ray activity for the two years is apparent.
\\
Figure \ref{crab} shows that our experiment was stable. We choose to interpret the excess as due to gamma-rays 
from this well-known emitter. 
We use the same acceptances as for Mrk~421, again matched to the conditions of each data run. 
The flux at 110 GeV thus obtained for the year 2000 is $(0.5\pm 0.15)\times 10^{-10}~\mathrm{erg\,cm^{-2}\,s^{-1}}$ 
(Statistical errors only). 
Combining both years decreases the result by less than $1\sigma$. 
Figure~\ref{spec501} shows the flux superimposed on an SED taken from (\cite{egret501}). 
The ASCA X-ray and EGRET gamma ray data were simultaneous with each other.
We have added the BeppoSAX X-ray data taken at the same time as CELESTE's.
The TeV spectrum for the 1998 season from the CAT Cherenkov imager 
corresponds to a quiet X-ray period (\cite {cat501bis}), 
with an average of $0.6$ ASM counts per second, close to the year 2000 rate, and no significant flares.

\subsection{1ES~1426+428} 

\begin{figure*}
\hspace*{-3.cm}
\centering\includegraphics[width=0.7\textwidth, angle=270]{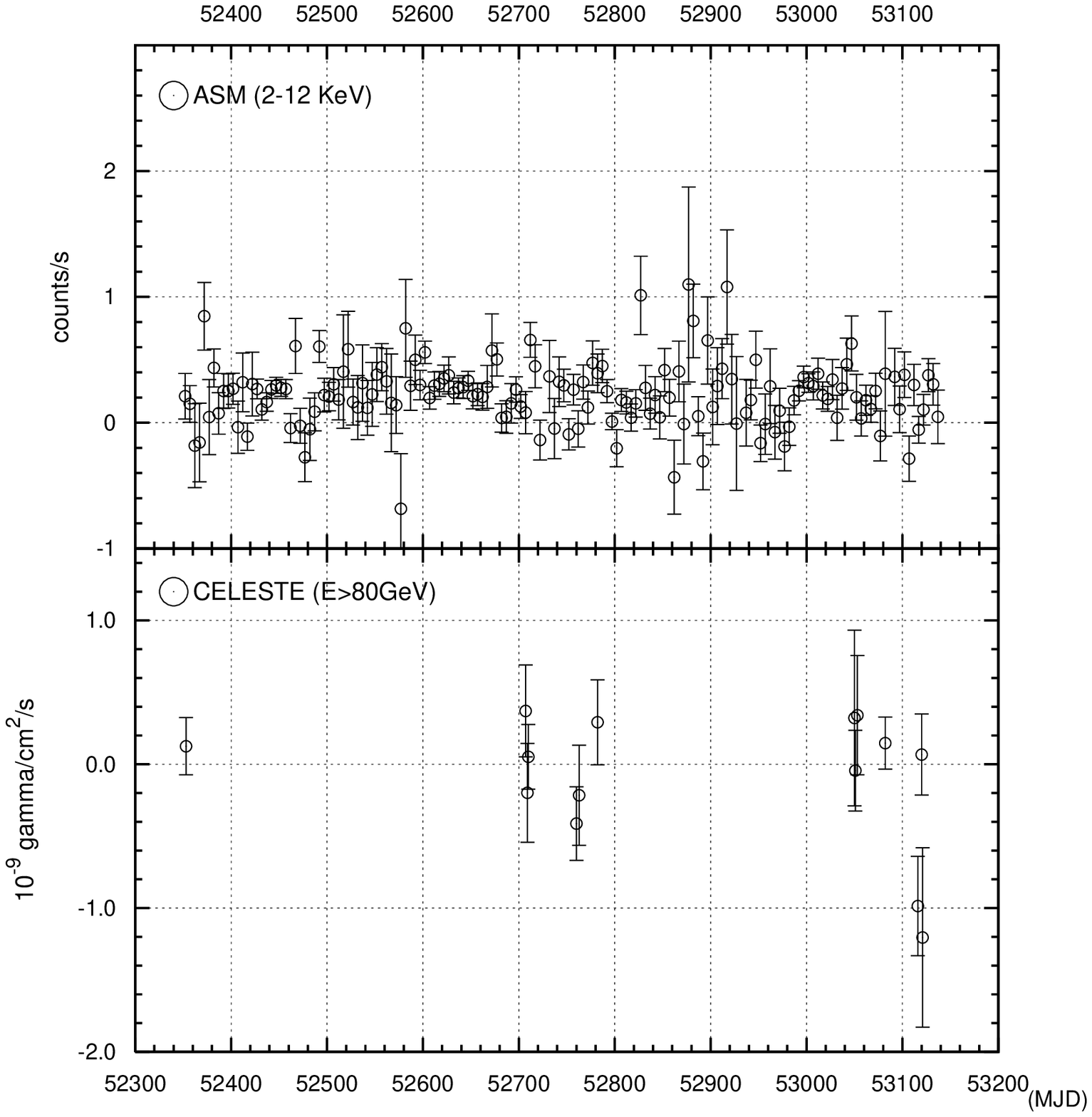}
\caption{Light curves for 1ES~1426+428~--  CELESTE detected no signal.
The CELESTE daily averages run from 2002 March 22 to 2004 April 25. The ASM points are 5 day running averages. 
   \label{1426lc}}
\end{figure*}

Table~\ref{results1426tab} and the light curve in
Figure~\ref{1426lc} show that CELESTE detected no gamma-rays from 1ES~1426+428, indicating
that the $2\sigma$ upper flux limit at 80 GeV is $0.2 \times 10^{-10}~\mathrm{erg\,cm^{-2}\,s^{-1}}$. 
The average ASM X-ray flux during CELESTE's observations was $0.2 \pm 0.1$ counts per second.
For the two seasons leading to the detection of this blazar with the Whipple telescope the ASM 
rate ranged from $0.15 \pm 0.04$ to $0.65 \pm 0.06$, with an average near $0.3$, with no discernible
correlation between the weak X-ray and TeV signals (\cite{whipple1426}).

Prior to the recent HESS result in (\cite{hessIR}), 1ES~1426+428 was the most distant blazar 
used to constrain the density of extragalactic diffuse infrared
light, possible because TeV photons can be absorbed by low energy photons via
electron-positron pair production, $\gamma \gamma \rightarrow e^+e^-$. In particular,
(\cite{ir}) took the spectra observed at high energies and, for a range of plausible
assumptions about infrared spectral shapes and densities, removed absorption effects.
Extrapolations of the de-absorbed spectra to 100 GeV suggested fluxes detectable by CELESTE. 
In the end, our non-detection during a low X-ray state does not constrain the diffuse infrared densities.

\begin{table*}
\begin{center}
%\resizebox{\hsize}{!}{
\begin{tabular}{llllll}
 &  & \multicolumn{3}{c}{Number of events} & Significance\\
Data set &Cut & $N_\mathrm{\textit{\scriptsize{On}}}$ & $N_\mathrm{\textit{\scriptsize{Off}}}$ & $N_\mathrm{\textit{\scriptsize{On}}}-N_\mathrm{\textit{\scriptsize{Off}}}$ & $N_\sigma$\\
\hline
All data (SPV, $8.8~\mathrm{h}$) & Raw data & $675\,340$ & $673\,239$ & $2\,101$ &  \\
& Software trigger & $455\,794$ & $453\,660$ & $2\,134$ & $2.0$ \\
& All cuts & $64\,230$ & $64\,433$ & $-203$ & $-0.5$\\
\end{tabular}
%}
\caption{Number of events before and after cuts for the 1ES~1426+428 data sets.
$N_\sigma$ is calculated after correcting for the $\sim 80\%$ data acquisition
efficiency.
}
\label{results1426tab}
\end{center}
\end{table*}

\section{Conclusions} 
We have provided some of the first astrophysical flux measurements around 100~GeV, beyond the reach of EGRET on the
Compton CGRO and below the range of previous ground-based telescopes, by recording atmospheric Cherenkov light with 
the reconverted solar tower facility at Th\'emis. 
We have increased our detector sensitivity as compared to our earlier work, to a level near
that predicted in our original proposal. The gain came mainly from a better analysis procedure
but also from refinements in our detector simulation and electronics improvements. 80~\% of
the gamma-rays we detect have energies between 50 and 350~GeV. Bad weather at our site lead to
a low duty-cycle, resulting in a modest number of detections in spite of our detector's good
performance.

Our results for the blazar Mrk~421 help constrain the shape of the high energy peak of the spectral
energy distribution. We have made the first detection of Mrk~501 in this range, although the detection 
is weak, and we have placed an upper limit on the flux from the distant blazar 
1ES~1426+428 during a period of X-ray inactivity.

\begin{acknowledgements} This work was supported by the IN2P3 of the French National Center for Scientific Research
(CNRS) and by the R\'egion Languedoc-Roussillon. We are grateful for the cooperation of Electricit\'e de France. 
Antoine Perez, Jacques Maurand, and St\'ephane Rivoire kept the Th\'emis facility operational through hard work
and dedication.
We thank Wei Cui, Jun Kataoka, and Dieter Horns for providing us with data files and macros.  \end{acknowledgements}


\begin{thebibliography}{}

\bibitem[Aharonian et al. 2002] {hegra421a} Aharonian, F.A. et al., A\&A 393, 89-90 (2002)
\bibitem[Aharonian et al. 2003] {hegra421b} Aharonian, F.A. et al., A\&A 410, 813-821 (2003)
%\bibitem[Aharonian et al. 2003b] {hegra1959} Aharonian, F.A. et al., A\&A 406, L9-L13 (2003)
\bibitem[Aharonian et al. 2006] {hessIR} Aharonian, F.A. et al., Nature 440, 1018 (2006)
\bibitem[Aharonian et al. 2004] {hegraCrab} Aharonian, F.A. et al., ApJ 614, 897-913 (2004)
\bibitem[see ASM in references] {ASM} ASM (All Sky Monitor) quick-look results are provided by the ASM/RXTE team,
http://xte.mit.edu/ASM\_lc.html
\bibitem[Baillon et al. 1993] {themistocle} Baillon, P. et al., Astropart. Phys. 1, 341-355 (1993)
\bibitem[Bernl\"ohr 2000] {bernlohr} Bernl\"ohr, K., Astropart. Phys. 12, 255-268 (2000)
\bibitem[B{\l}a{\.z}ejowski et al. 2005] {blaze} B{\l}a{\.z}ejowski, M. et al., ApJ 630, 130-141 (2005)
\bibitem[Brion 2004] {EBcospar} Brion, E. et al, proc. $35^{th}$ COSPAR, Paris (2004). 
\bibitem[Brion 2005] {EBthesis} Brion, E., doctoral thesis (2005), \mbox{http://doc.in2p3.fr/themis/CELESTE/PUB/papers.html} 
\bibitem[Bruel 2004] {philippe} Bruel, P., proc. ``Physics \& Astrophysics in Space'', Frascati (2004)
\bibitem[Boone et al. 2002] {stacee421} Boone et al., ApJL 579, 5-8 (2002)
\bibitem[Carson 2005] {CarsonThesis} Carson, J., doctoral thesis, \mbox{http://www.astro.ucla.edu/\~{}carson} (2005), ApJ in preparation
%\bibitem[Catanese et al. 1998] {whipple2344} Catanese, M. et al., ApJ 501, 616 (1998)
%\bibitem[Cui 2004] {Cui} Cui, W., ApJ 605, 662-669 (2004)
\bibitem[Djannati-Ata{\"{\i}} et al. 1997] {cat501} Djannati-Ata{\"{\i}}, A. et al., A\&A 350, 17-23 (1997)
\bibitem[Djannati-Ata{\"{\i}} et al. 2002] {cat1426} Djannati-Ata{\"{\i}}, A. et al., A\&A 391, L25-L28 (2002)
\bibitem[Dwek \& Krennrich 2005] {ir} Dwek, E. \& Krennrich, F., ApJ 618, 657-674 (2005)
\bibitem[Gehrels \& Michelson 1999] {glast} Gehrels, N. and Michelson, P., Astropart. Phys. 11, 277-282 (1999) 
\bibitem[Giebels et al. 1998] {bgnim} Giebels, B. et al., Nucl. Instr. Meth. A 412, 329-341 (1998)
\bibitem[Goret et al. 1993] {asgat} Goret, P. et al., A\&A 270, 401-406 (1993)
\bibitem[Hartman et al. 1999] {3rdcatalog} Hartman, R.C. et al., ApJS 123, 79-202 (1999)
\bibitem[Hayes \& Latham 1975] {HL} Hayes, D.~S. and Latham, D.~W., ApJ 197, 593-601 (1975)
\bibitem[Heck et al. 1998] {corsika} Heck, D. et al., Report FZKAS 6019, Forschungszentrum Karlsruhe (1998)
\bibitem[Holder et al. 2001] {jamie} Holder, J. et al., 27th ICRC, Hamburg (2001)
\bibitem[Horan et al. 2002] {whipple1426} Horan, D. et al., ApJ 571, 753-762 (2002)
%\bibitem[Holder et al. 2003] {whipple1959} Holder, J. et al., ApJL 583, 9-12 (2003)
\bibitem[Kataoka et al. 1999] {egret501} Kataoka, J. et al., ApJ 514, 138-147 (1999)
\bibitem[Katarzy\'nski et al. 2001] {KSK} Katarzy\'nski, K., Sol, H. and Kus, A., A\&A 367, 809-825 (2001)
\bibitem[Kertzman \& Sembroski 1994] {kaskade} Kertzman, M.~P. and Sembroski, G.~H., Nucl. Instrum. Methods A 343, 629-643 (1994)
%\bibitem[Khelifi et al. 2001] {bruno} Khelifi, B. et al. 27th ICRC, Hamburg (2001) % This is '421
\bibitem[Krawczynski et al. 2001] {feb2000} Krawczynski et al., ApJ 559, 187-195 (2001)
\bibitem[Krawczynski et al. 2004] {kwaz1959} Krawczynski et al., ApJ 601, 151-164 (2004)
%\bibitem[Krennrich et al. 2001] {whipple421501} Krennrich, F. et al., ApJL 560, 45-48 (2001)
\bibitem[Lavalle et al. 2006] {M31} Lavalle, J., Manseri, H., et al., A\&A 450 1-8 (2006)
%\bibitem[Manseri 2004] {HMthesis} Manseri, H., ``Astronomie gamma au dessus de 30~GeV - une nouvelle m\'ethode d'identification des rayons $\gamma$ cosmiques \`a partir du sol avec le d\'etecteur CELESTE'', doctoral thesis, Universit\'e Paris VI (2004), \mbox{http://doc.in2p3.fr/themis/CELESTE/PUB/papers.html}
\bibitem[Manseri 2004] {HMthesis} Manseri, H., doctoral thesis (2004), \mbox{http://doc.in2p3.fr/themis/CELESTE/PUB/papers.html}
\bibitem[Massaro et al. 2004] {beppo} Massaro, E. et al.,  A\&A 422, 103-111 (2004)
%\bibitem[Massaro et al. 2006] {massaroIII} Massaro, E. et al.,  A\&A submitted, astro-ph/0511673
\bibitem[Mukherjee et al. 1997] {egretblazars} Mukherjee, R. et al., ApJ 490, 116 (1997)
%\bibitem[de Naurois 2000] {thesismdn} de Naurois, M., ``Reconversion d'une centrale solaire pour l'astronomie $\gamma$. Premi\`ere observation de la N\'ebuleuse du Crabe et du Blazar Markarian~421 entre $30$ et $300\,\mathrm{GeV}$, doctoral thesis, Universit\'e Paris VI (2000), \mbox{http://polywww.in2p3.fr/celeste/these/These.html}
%\bibitem[de Naurois 2000] {thesismdn} de Naurois, M., doctoral thesis (2000), \mbox{http://polywww.in2p3.fr/celeste/these/These.html}
\bibitem[de Naurois et al. 2002] {ourcrab} de Naurois, M., Holder, J. et al., ApJ 566, 343-357 (2002)
\bibitem[Oser et al. 2001] {staceecrab} Oser, S. et al., ApJ 547, 949-958 (2001)
%\bibitem[Perlman et al. 1996] {Perlman} Perlman, E.~S. et al., ApJS 104, 251 (1996)
%\bibitem[Petry et al. 2002] {whipple1426} Petry, D. et al., ApJ 580, 104-109 (2002)
\bibitem[Piron et al. 2001a] {cat421} Piron, F. et al., A\&A 374, 895-906 (2001a)
\bibitem[Piron et al. 2001b] {cat501bis} Piron, F. in proc. SF2A-Lyon astro-ph/0109286 (2001b)
\bibitem[Piron et al. 2003] {fred} Piron, F. et al., 28th ICRC, Universal Academy Press, vol. 5 p.2607 OG 2.3 (Tsukuba 2003)
% above is fred's spectrum, with 4 points.
\bibitem[Par\'e et al. 2002] {repoman} Par\'e, E. et al., Nucl. Instr. Meth. A 490, 71-89 (2002)
%\bibitem[Schroedter et al. 2005] {schroedter} Schroedter, M. et al., ApJ 628, 617-628 (2005)
%\bibitem[Smith et al. 1996] {proposal} Smith, D.~A. et al., CELESTE experimental proposal (1996), \mbox{http://polywww.in2p3.fr/celeste/public/cxp.ps.gz}
%\bibitem[Smith 2001] {dasmoriond} Smith, D.~A., proc. ``Very High Energy Phenomena in the Universe'', Moriond (2001)
%\bibitem[Smith 1998] {Smith98} Smith, D.~A., ``First Detection of Gamma Rays from the Crab Nebula with the CELESTE Solar Farm Cherenkov Detector'', plenary Highlight talk in proc. of the 19th Texas Symposium on Ultrarelativistic Astrophysics, Paris (1998)
\bibitem[Smith 1998] {Smith98} Smith, D.~A., proc. 19th Texas Symposium on Ultrarelativistic Astrophysics, Paris (1998)
%\bibitem[Smith 2000] {Smith00} Smith, D.~A., Nucl. Phys. 80B, 163-172 (2000)
%\bibitem[Smith 2004] {note53} Smith, D.~A., CELESTE internal note 53 (2004)
\bibitem[Smith 2005] {Smith05} Smith, D.~A., proc. ``Atmospheric Cherenkov Detectors VII'', Palaiseau (2005)
\bibitem[Tavecchio et al. 1998] {TMG} Tavecchio, F., Maraschi, L. and Ghisellini, G., ApJ 509, 608-619 (1998)
%\bibitem[Tavecchio et al. 2001] {tavecchio} Tavecchio, F. et al., ApJ 554, 725-733 (2001)
%\bibitem[Xue \& Cui 2005] {XueCui} Xue, Y. and Cui, W., ApJ 622, 160-167 (2005)
\end{thebibliography}
\end{document}